\documentclass[pre,twocolumn,preprintnumbers,amsmath,amssymb,floatfix]{revtex4}

\usepackage{amsmath}
\usepackage{amsfonts}
\usepackage{bbold}
\usepackage{mathrsfs}
\usepackage{color}
\usepackage{epstopdf}
\usepackage{graphicx}
\usepackage{bm}
\usepackage{silence}
\usepackage{amssymb}
\usepackage{bm}
\usepackage{graphicx}
\usepackage{color}
\usepackage{amsfonts}
\usepackage{subfigure}

\def\ovm#1{\bgroup \color{blue} OVM: #1\egroup}

\newcommand{\rr}{{\mathbf r}}
\newcommand{\f}{{\mathbf f}}

\newcommand{\e}{{\hat{\mathbf e}}}

\newcommand{\matteo}[1]{{\color{black} #1}}

\newcommand{\BEQ}{\begin{equation}}
\newcommand{\EEQ}{\end{equation}}
\newcommand{\BEA}{\begin{eqnarray}}
\newcommand{\EEA}{\end{eqnarray}}

\begin{document}
\title{Narrow-escape time and sorting of active particles in circular domains}

\author{Matteo Paoluzzi \footnote{Present address: Departament de Física de la Mat\`eria Condensada, Universitat de Barcelona, C. Martí Franqu\`es 1, 08028 Barcelona, Spain.} }
\email{mttpaoluzzi@gmail.com}

\author{Luca Angelani}
\author{Andrea Puglisi}


\affiliation{
ISC-CNR,  Institute  for  Complex  Systems and 
Dipartimento di Fisica, Sapienza Universit\`a di Roma, Piazzale A. Moro 2, I-00185, Rome, Italy 
}

\begin{abstract}
It is now well established that microswimmers can be sorted or
segregated fabricating suitable microfluidic devices or using external
fields. 
A natural question is how these techniques can be employed
for dividing swimmers of different motility.
In this paper, using
numerical simulations in the dilute limit, we investigate how motility
parameters (time of persistence and velocity) impacts the
narrow-escape time of active particles from circular domains. 
 We show that the escape time undergoes a crossover
between two asymptotic regimes. The control parameters of the crossover
is the ratio between persistence length of the active motion and the typical length scale of the circular domain.
We explore the
possibility of taking advantage of this finding for sorting active
particles by motility parameters.
\end{abstract}

\maketitle

\section{Introduction}\label{Introduction}

Active particles are widespread in nature
\cite{Bechinger17,Marchetti13}.  Because of their autonomous motion,
active particles break fluctuation-dissipation theorem at
single-particle level \matteo{\cite{maggi2017memory}} making possible a rich phenomenology that does
not share any similarity with equilibrium systems
\cite{zhang2010collective,Lushi14,bricard2013emergence,sanchez2012spontaneous}.
%
During the last decades, it has been shown that active particles and
swimming organisms can be employed for actuating micro-motors
\cite{angelani2009self,di2010bacterial,Sokolov10,maggi2016self}
, controlling and stabilizing density fluctuations
\cite{Galajda07,bricard2013emergence}, or for driving macroscopic
directed motion \cite{Angelani11,Reichhardt17}.
Many aspects of this remarkably phenomenology can be rationalized
starting from the morphological properties of the single-particle
trajectory.  The typical trajectory of an active particle is well
captured by a persistent random walk.  In particular, the existence of
a finite persistence length gives rise to a motion that is ballistic
on a short-time scale and it becomes diffusive for larger times.  It
is now well established that, because of the finite persistency,
active particles slow down in regions where they are denser
\cite{schnitzer1993theory} and accumulate at the boundaries of a
confining container \cite{VladescuPRL2014,das2018confined}.
Remarkably, simple artificial environments can be designed for sorting
active particles in small regions of space \cite{Galajda07}.
However, sorting particles dynamically from slower to faster remains a
challenge \matteo{\cite{mijalkov2013sorting}}. 
While some attempts to obtain particles segregation
 have been made by using external fields \matteo{\cite{PhysRevLett.108.268307,costanzo2014motility,PhysRevE.99.032605},} 
designing machinery suitable for segregating particles of
different motility properties without using any external potential
could have important applications, as in the case of in vitro
fertilization where the identification and gathering of motile sperms
without invasive techniques is a hard task
\cite{nosrati2017microfluidics,koh2015study,gaffney2011mammalian,campana1996intrauterine}.

In this work, we will focus our attention on the narrow-escape problem
of active particles \cite{benichou2014first,singer2006narrowII}.  We
are interested in studying this problem numerically in two dimensions
considering a circular container with a small target exit site on the
boundary. The target site allows particles to escape from the
confining structure and we assume that the particles cannot come back
into the chamber.
Looking at the properties of
the first-passage time for a particle to escape from the chamber, we
compare two paradigmatic active dynamics, i. e., Run-and-Tumble and
Active Brownian.  In agreement with recent studies on optimal search
strategies with active particles \cite{PhysRevE.94.012117}, both the
active dynamics show a crossover between two regimes in the mean
first-passage time.  The first regime, typical of active systems,
takes place when the persistence length of the random walk is larger
than the size of the confining structure.  The second regime is
reached in the diffusive limit, i. e., when the persistence length is
small when compared with the size of the chamber.

Our findings show that, although both dynamics show exactly the same
diffusive limit, in the active limit the comparison of different
active dynamics at equal persistence times show up differences, with
Run-and-Tumble particles being less efficient than Active Brownian in
escaping from the chamber.  We identify an empirical function
$f(x)$ that matches smoothly the two regimes, with $x=\ell / R$, being
$\ell$ the persistence length of the active motion and $R$ the radius
of the confining structure. The function captures the crossover
between the two scaling regimes that takes place for $x \approx 1$.

Since the two regimes can be reached in either way, by varying the
motility parameters or by tuning the size of the confining structure,
we show that one can take advantage of the crossover between active
and diffusive regime for sorting particles of different persistence
length by varying the size of the structure.

\section{Model and Methods} \label{Model}

\begin{figure}[!t]
\centering
\includegraphics[width=.5\textwidth]{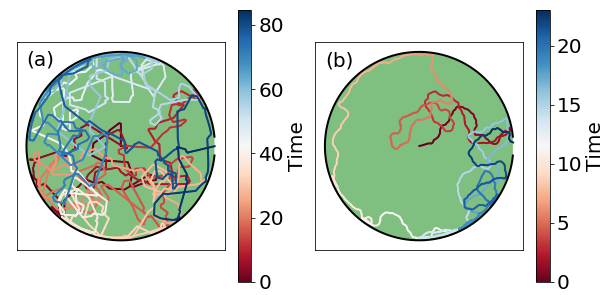} 
\caption{ Active particles escaping from circular domains. A
  representative trajectory for a single Run-and-Tumble particle 
    (a) and a single Active Brownian particle (b). \matteo{The parameters are $\lambda=D_r=1,v_{self}=1,R=10 (2\sigma)\matteo{=10}$} Colors
  change from red to blue as time increases. In both cases the
  particle is injected at the center of the chamber and escapes from
  the circular domain, reaching a slit of width $\delta$ symmetrically
  displaced around $(R,0)$. 
}
\label{fig:fig0}      
\end{figure}

As a model system, we consider a gas of $N$ non-interacting active
particles in two spatial dimensions
confined to stay inside a circular chamber of radius
$R$. The chamber has a slit of size $\delta$ 
on the boundary where particles can
escape (they never come back).  Indicating with $O=(0,0)$ the center
of the chamber, the slit is displaced symmetrically around $(R,0)$,
i. e., with coordinates $(R \cos \varphi, R \sin \varphi)$ and $(R
\cos \varphi, -R \sin \varphi)$, with \matteo{$\varphi=\delta / 2 R$.} 
    In the present work, we present
results for a fixed value of \matteo{$\delta = 2 \sigma$}, with $\sigma$ the
particle radius~\cite{Paoluzzi15}, see below. 

\begin{figure*}[!t]
\centering
\includegraphics[width=1.\textwidth]{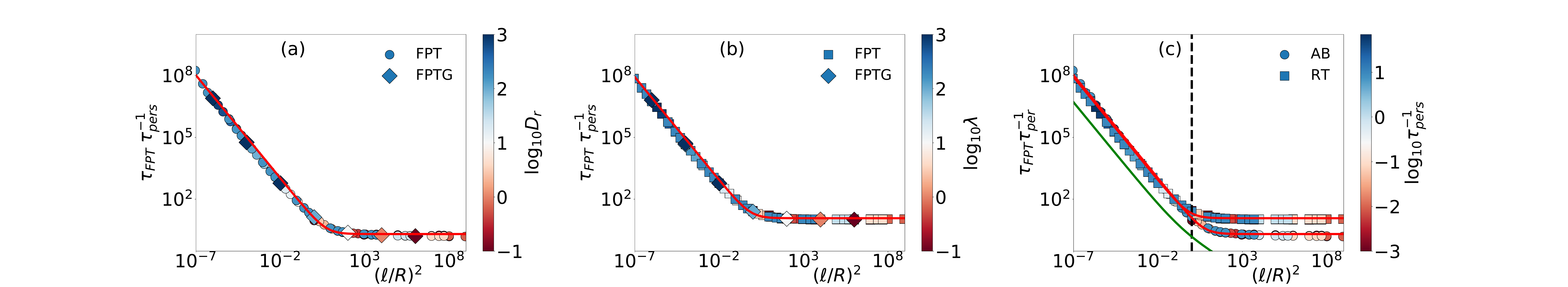} 
\caption{ Mean First-Passage Time of active walkers escaping from
  circular domains.  $\tau_{FPT}$, in unit of persistence time
  $\tau_{pers}$ (which is $1/\lambda$ or $1/D_r$ for Run-and-Tumble or
  Active Brownian particles respectively), is reported against the
  nondimensional parameter $\ell^2 / R^2$, for Active Brownian (a)
  and Run-and-Tumble (b) particles. Square (diamonds) indicate results
  with an initial position set at the center of the chamber (uniformly
  distributed in the chamber). (c) Comparison between Run-and-Tumble
  and Active Brownian dynamics. 
  \matteo{The persistence length is varied by exploring different motility parameters, i. e.,
  $\tau_{pers} \in [10^{-3},10]$, $v_{self} \in[10^{-2},10^{3}]$, and also by changing the chamber size, i. e., $R \in [5,8\times10^2]$.}
  The red lines are fit to the
  empirical function $f(x)$, see text, Eq. (3). The green curve in (c)
  is Eq. (4).
}
      \label{fig:fig1}
\end{figure*}

\begin{figure}[!t]
\centering
\includegraphics[width=.5\textwidth]{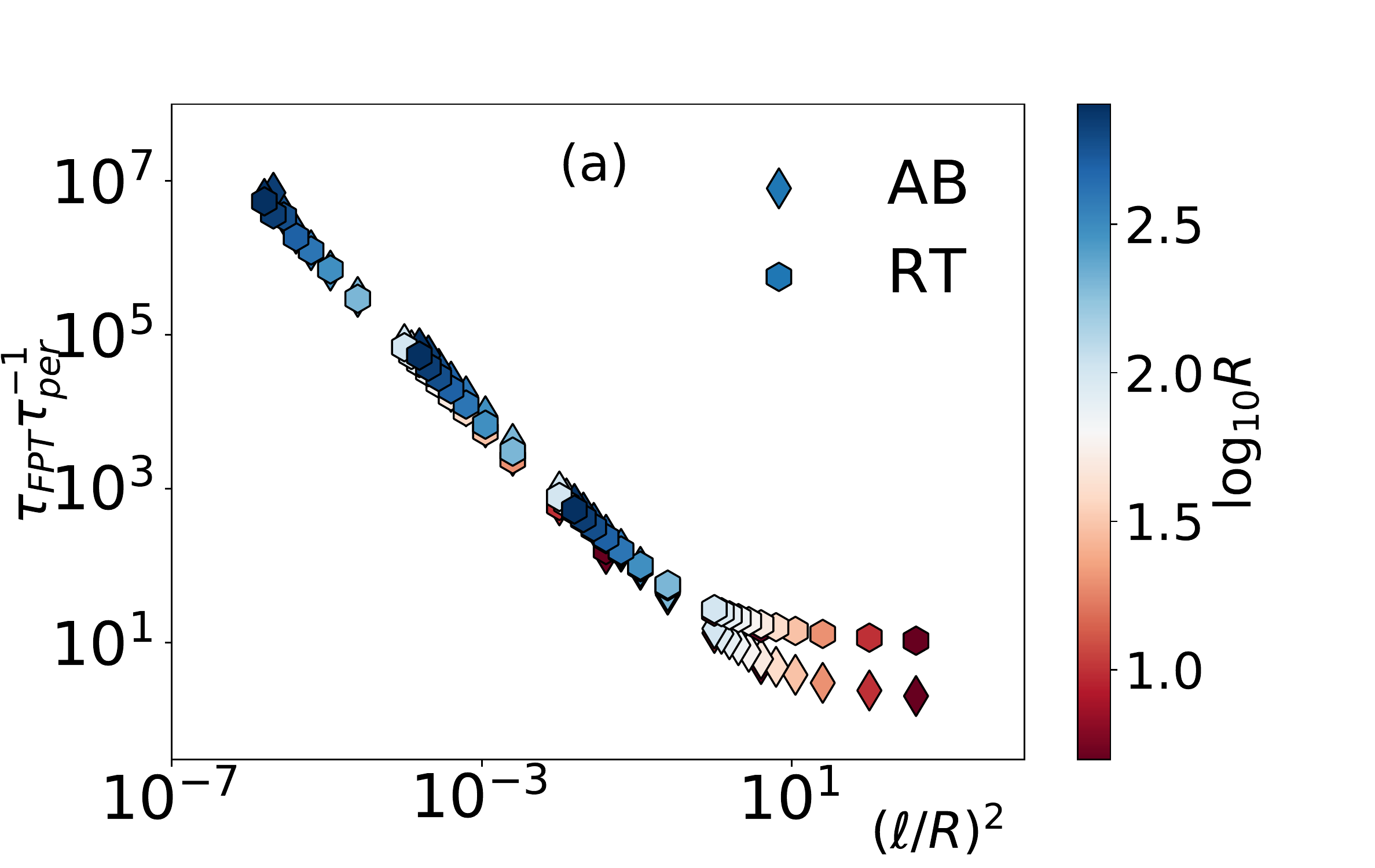} \\
\includegraphics[width=.5\textwidth]{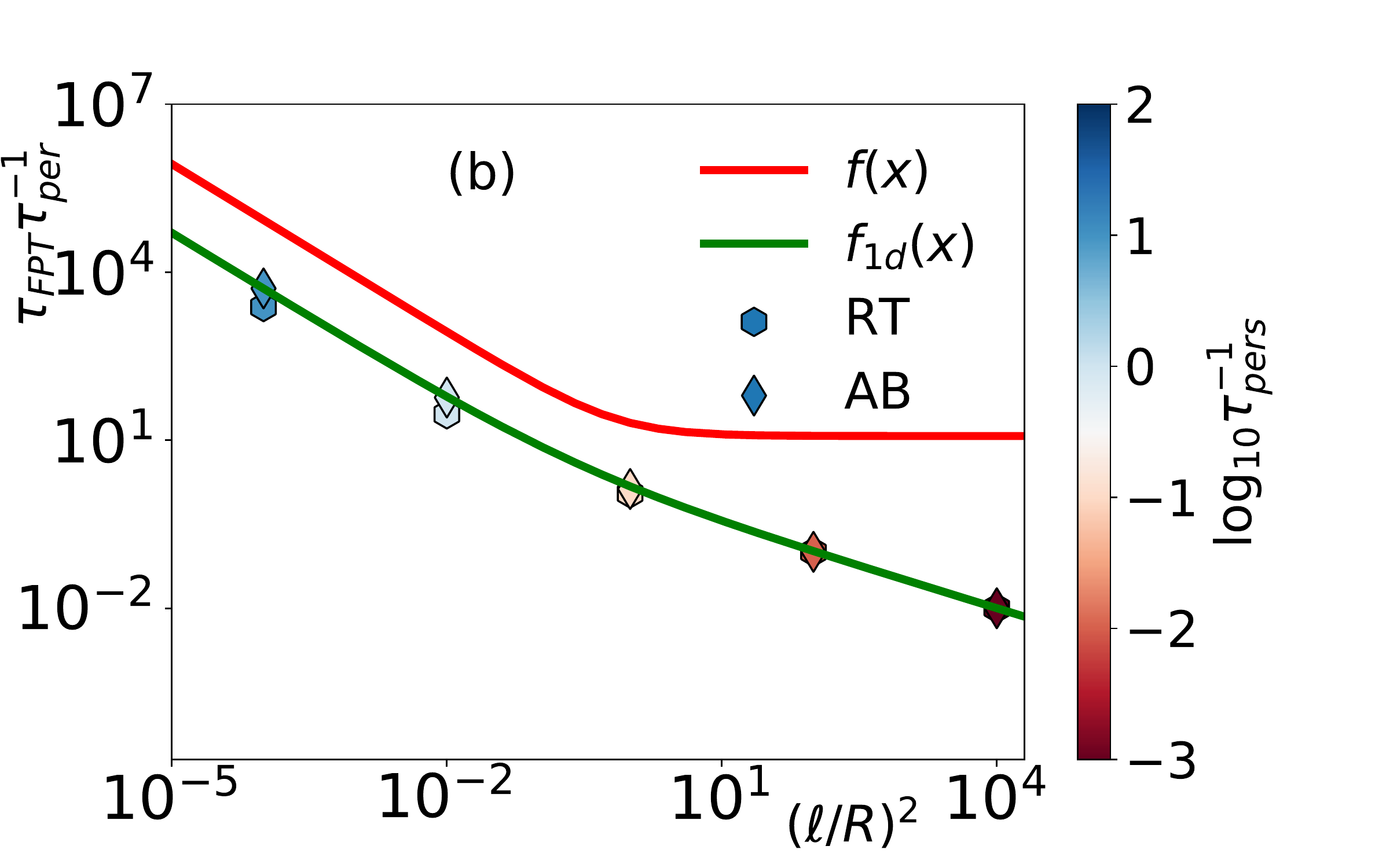} 
\caption{ \matteo{(a) Mean First-Passage Time as a function of the nondimnesional $\ell^2 / R^2$ 
for different chamber sizes ranging from $R=5$ to $R=8 \times 10^2$. (b) Mean First-Passage Time of reaching the boundary, the green curve is
Eq. (3), for comparison we include Eq. (4) as a guide to the eye (red curve).
}}
      \label{fig:fig1b}
\end{figure}

The confining structure is implemented through
the image technique, i. e., the active particle sees its own image
located at \matteo{$R+ 2 \sigma$, this choice ensures that the particle is point-like and it is confined to displace at distances smaller or equal to $R$} \cite{Paoluzzi15}.  Each particle does not
interact with the others.  
\matteo{It is worth noting that the active walkers are point-like. However, since we are modeling the mechanical interaction
with the confining walls using the image technique, the distance between the point-like particle and the force center of its image naturally introduce
the length scale $\sigma$.}
\matteo{In the following, we express length 
in unit of $2\sigma$.}
For the active dynamics, we consider two
microscopic models: Run-and-Tumble and Active Brownian.
Run-and-Tumble dynamics has been implemented following
Refs. \cite{angelani2009self,paoluzzi2014run,paoluzzi2013effective,paoluzzi2018fractal}.
Considering the case of overdamped dynamics, that is a good approximation at low Reynolds number, 
the equation of motion for the particle $i$ is
\BEQ \label{dyn}
 \dot{\rr}_i(t) = v_{self} \e_i + \mu \f_{image} \; .
\EEQ
The versor $\mathbf{e}_i=(\cos \theta_i, \sin \theta_i)$ specifies the
swimming direction. $\mu=1$ is the mobility, $\f_{image}$ is the
short-range force exerted by the image, i. e., $\f_{image} = -\nabla
\phi(r_{image})$ with $\phi(r)=A r^{-12}$
\cite{angelani2009self}.  
  The evolution of $\e_i$
depends on the model we consider. In the case of Run-and-Tumble
dynamics \cite{schnitzer1993theory}, $\e_i$ stochastically rotates
with a rate $\lambda$, i. e., the tumbling rate. Meaning that a new
orientational angle $\theta$ is extracted by a uniform distribution
in $[0,2\pi]$ 
and then it remains
constant for a time that is poissonian distributed with rate $\lambda$.
%
For Active Brownian
particles \cite{das2018confined}, the angle $\theta_i$ undergoes the
following Langevin dynamics
\BEQ
 \dot{\theta}_i(t) = \eta_i
\EEQ
with $\langle \eta_i(t) \rangle = 0$ and $\langle \eta_i(t) \eta_j(s)
= 2 D_r \delta_{ij} \delta(t-s)$, with $D_r$ the rotational diffusion
constant.

The tumbling rate $\lambda$ and the diffusion coefficient $D_r$ fix
$\tau_{pers}$ that is $\tau_{pers}=\lambda^{-1}$,
and $\tau_{pers}=D_r^{-1}$, for the two active dynamics, respectively.
Using the self-propulsion velocity $v_{self}$, we can define the
persistence length $\ell = v_{self} \tau_{pers}$.  In both cases, the
motion is characterized by a ballistic regime on times $t \ll
\tau_{pers}$ and a diffusive regime for $t \gg \tau_{pers}$.  In two
spatial dimensions, the diffusion can be described through the
effective diffusion coefficient $D_{eff} = v_{self}^2 \tau_{pers} /
2$.
\matteo{We explore a wide range of persistence length by varying motility
parameters $\tau_{pers} \in [10^{-3},10]$, $v_{self} \in[10^{-2},10^{3}]$. 
The chamber size is changed within the interval $R \in [5,8\times10^2]$.}

We solve Eq. (\ref{dyn}) numerically integrating the equation of
motion \matteo{using Euler method with a time step $\Delta t$ ranging from
$10^{-3}$ to $10^{-5}$.} 
    \matteo{The numerical integration is performed} until each of the runners has reached the target.
In the first part of our work, we consider a gas of non-interacting 
active particles
with same motility parameters $v_{self}$ and $\tau_{pers}$.  Particles
are injected in the center of the chamber at $t=0$ and thus the
initial density profile $n(\rr,0)$ reads $n(\rr,0) = \frac{N}{\pi R^2}
\delta(\rr)$. We also consider the situation where particles are
uniformly distributed at $t=0$.  With both kinds of initial conditions
we evaluate the Mean First-Passage Time $\tau_{FPT}$ defined as the
average time required for escaping from the chamber.  $\tau_{FPT}$ is
computed considering the escape of $N=10^5$ particles.

In the second part we consider a mixture of particles injected in the
center, with motility parameters ($v_{self}$ or $\tau_{pers}$)
extracted from a uniform distribution. In this case, we are interested
in the evolution of $n(v,t)$ and $n(D_r,t)$ with $n(.,t) \equiv N(.,t)
/ N(.,0)$ and $N(.,t)$ the number of particles at time $t$ with a
given value of motility parameters.

\section{Escape of a population with identical parameters} \label{results}

The typical trajectories of Run-and-Tumble and Active Brownian
dynamics are shown in Fig. (\ref{fig:fig0}). Both dynamics share the
same motility parameters, i. e., $\lambda=D_r=1$ and
$v_{self}=1$. Particles are confined into a circular chamber of radius
$R=10 (2\sigma)\matteo{=10}$.  As one can see, in either cases, the typical
trajectory is a persistent random walk, as it is well known in the
literature \cite{Bechinger17}. Active Brownian dynamics generate
smoother trajectories compared with those obtained through
Run-and-Tumble. The latter are characterized by straight run
interrupted by tumbling events.  Because of the confinement, one
expects to observe different dynamical behaviors depending on the
characteristic size of the circular chamber. In particular, we define
the {\it active regime} when the radius $R$ is smaller than the
persistence length $\ell$. We thus identify the opposite situation as
the {\it diffusive regime}, i. e., when $R \gg \ell$.

It is worth noting that a given value of $D_{eff}$ can be obtained
through different combinations of $v_{self}$ and
$\tau_{pers}$. Moreover, the genuine diffusive limit is recovered
performing simultaneously the limit $v_{self}\to \infty $ and
$\tau_{pers} \to 0$ at fixed $D_{eff}$, as it was realized by Kac in a
seminal paper on the telegrapher's equation \cite{kac1974stochastic}.
As we will see in the next section, both models show the same
diffusive limit that is consistent with the dynamics of a Brownian
walker coupled to a thermal bath with effective temperature $T_{eff} =
\mu D_{eff}$. To conclude this overview of model parameters, we recall
that 
\matteo{$2 D_{eff}\tau_{pers}/R^2 = \ell^2/R^2$,}
therefore fixing (at given
$R$) $\tau_{pers}$ and $\ell$ also fixes $D_{eff}$. Most importantly,
the condition $R \ll \ell$ ($R \gg \ell$) is equivalent to
$D_{eff}\tau_{pers} \gg R^2$ ($D_{eff}\tau_{pers} \ll R^2$).

Also in the {\it active regime}, $\tau_{FPT}$ follows the same
asymptotic scaling behavior in both models.  However, the fact that
Run-and-Tumble and Active Brownian trajectories are {\it
  morphologically} different has a quantitative impact on $\tau_{FPT}$
away from the diffusive limit. This result is consistent with escape
time of active particles from a maze \cite{khatami2016active}.



The mean first-passage time $\tau_{FPT}$ is shown in
Fig. (\ref{fig:fig1}), rescaled by the persistence time
$\tau_{pers}$. Following the previous discussion, as non-dimensional
control parameter we use $(\ell / R)^2 \equiv \matteo{2} D_{eff}\tau_{pers}/R^2$.
Here we are considering a
situation where the persistence length is changed by varying both the
motility parameters, i. e., the persistence time $\tau_{pers}$, and
the self-propulsion velocity $v_{self}$ (therefore different values of
$D_{eff}$ are considered).  Panel (a) referes to Active Brownian, panel
(b) to Run-and-Tumble.
We have also varied the radius $R$ of the chamber, see Panel (c).

The result is a collapse of data onto a master curve which is similar
but not identical for the two dynamics.  In both dynamics
$\tau_{FPT}/\tau_{pers}$ undergoes a crossover from large values at
small persistence length (diffusive regime) to small values at large
persistence length (active regime).  The color code indicates the
inverse of the values of the persistence time $\tau_{pers}$.  We have
also reported data obtained with a different initial condition
(diamonds in figure).  In this case, the starting position is
uniformly distributed and the results neatly superimpose on the master
curve.

On a more quantitative level, we see that the diffusive regime, $\ell \ll R$
($D_{eff}\tau_{pers} \ll R^2$), is signaled by a scaling
$\tau_{FPT}/\tau_{pers} \sim (\ell/R)^{-2} = R^2/(D_{eff}
\tau_{pers})$ which implies $\tau_{FPT} \sim R^2/D_{eff}$.  In the
opposite limit, i. e., $\ell \gg R$, where the active
regime dominates, we observe a scaling $\tau_{FPT} \sim 
\tau_{pers}$. 

In panel (c) of the same figure we compare the master curves for the
Run-and-Tumble and Active Brownian dynamics.  In this case, we are
varying only the persistence time $\tau_{pers}$, i. e., the tumbling
rate $\lambda$, in the case of Run-and-Tumble particles, and the
rotational diffusion $D_r$, for Active Brownian particles.  The {\it
  diffusion limit} is thus approached for $\tau_{pers} \to 0$, the
{\it active limit} as soon as $\tau_{pers} \to \infty$.  As one can
see, they reach exactly the same diffusive limit when $D_{eff} R^{-2}
\to 0$.  In the opposite limit, i. e., $D_{eff} R^{-2} \to \infty$,
$\tau_{FPT} \sim \tau_{pers}$, however, Run-and-Tumble particles are
systematically slower in finding the exit than Active Brownian
particles.  We can guess that this difference is due to the fact that
Active Brownian particles, at variance with Run-and-Tumble, smoothly
change their self-propulsion direction. 

The behavior observed in the active regime is consistent with optimal
research strategies of run-and-tumble in spherical confinement
\cite{PhysRevE.94.012117}. The divergence of $\tau_{FPT}$ with the
persistence time can be rationalized noticing that, when $\tau_{pers}
\to \infty$, only the walkers moving towards the right direction,
i. e., with $\theta \in [-\varphi,\varphi]$, can escape from the
chamber, therefore the escape becomes more and more difficult. However
analytical predictions of $\tau_{FPT}$ for active dynamics remains an
open problem
\cite{PhysRevE.94.012117,redner2001guide,benichou2014first,singer2006narrow,singer2006narrowII,rupprecht2015exit}.
For the diffusive regime ($R \gg \ell$) the boundary is so far that
active particles displace a distance $\langle \Delta x^2 \rangle =
D_{eff} t$ and thus the boundary is reached on a time scale $R^{2}
/D_{eff}$, as we observe.

Noticeably, these two asymptotic regimes match smoothly at $\ell \,
R^{-1} = 1$. Indicating with $x=\ell^2 R^{-2} $ the control
parameter, we thus propose the following empirical function \matteo{$f(x)$} that
results suitable for capturing the whole emerging phenomenology \matteo{of the nondimensional quantity $\tau_{FPT}/\tau_{pers}$}
\BEQ 
\label{empir} f(x) = \frac{\alpha}{x} + \beta, 
\EEQ with $\alpha$
and $\beta$ which in principle may depend upon $R$. For instance
in~\cite{PhysRevE.94.012117} (where the geometry has important
differences with our setup) $\alpha \sim \log(R/\delta)$ and $\beta
\sim R/\delta$. 
Here we find \matteo{$\alpha=10.3$} and \matteo{$\beta=2.0$} for
Active Brownian particles at \matteo{$R=10$} and \matteo{$\alpha=8.5$} and \matteo{$\beta=11.9$}
for Run-and-Tumble particles at \matteo{$R=10$}. 
\matteo{It is worth noting that, for the space of parameters explored here, the parameters $\alpha$ and $\beta$
  turns out to be almost independent of $R$, as it is shown in Fig. (\ref{fig:fig1b})-(a).
}
Just for comparison, in the one dimensional case one has the exact 
expression for the mean first-passage time \cite{angelani2014first,angelani2015run}
\BEQ \label{fpt_free_1d}
f_{1d}(x) = 1/2x + 1/\sqrt{x} \; .
\EEQ
However, we note that this expression is obtained considering a
particle which reaches for the first time the boundary points,
corresponding in our two-dimensional case to a particle that escapes
once reaching the circular boundary, i.e., $\delta = 2 \pi R$.
\matteo{As one can see in Fig. (\ref{fig:fig1b})-(b), where we
  computed the escape time for reaching the boundary, $f_{1d}(x)$
  reproduces the numerical data. } At large persistence lengths ($x
\gg 1$) the time to hit the boundary is dominated by the term $\sim
1/ \sqrt{x} \sim 1/v_{self}$, as expected for a purely ballistic
motion. This behavior is quite different from that seen at large $x$
in Fig. 2, i.e. for the narrow escape problem: curve $f_{1d}(x)$ is
plotted in Fig. 2c for comparison.

These findings show that particles moving in an environment
characterized by a length scale $R > \ell $ are dramatically
disadvantaged in finding the exit with respect to particles such that
$R < \ell$. As we will show in the next section, this observation can
be employed for the purpose of sorting particles with different
persistent lengths. 

\begin{figure}[!t]
\centering
\includegraphics[width=.5\textwidth]{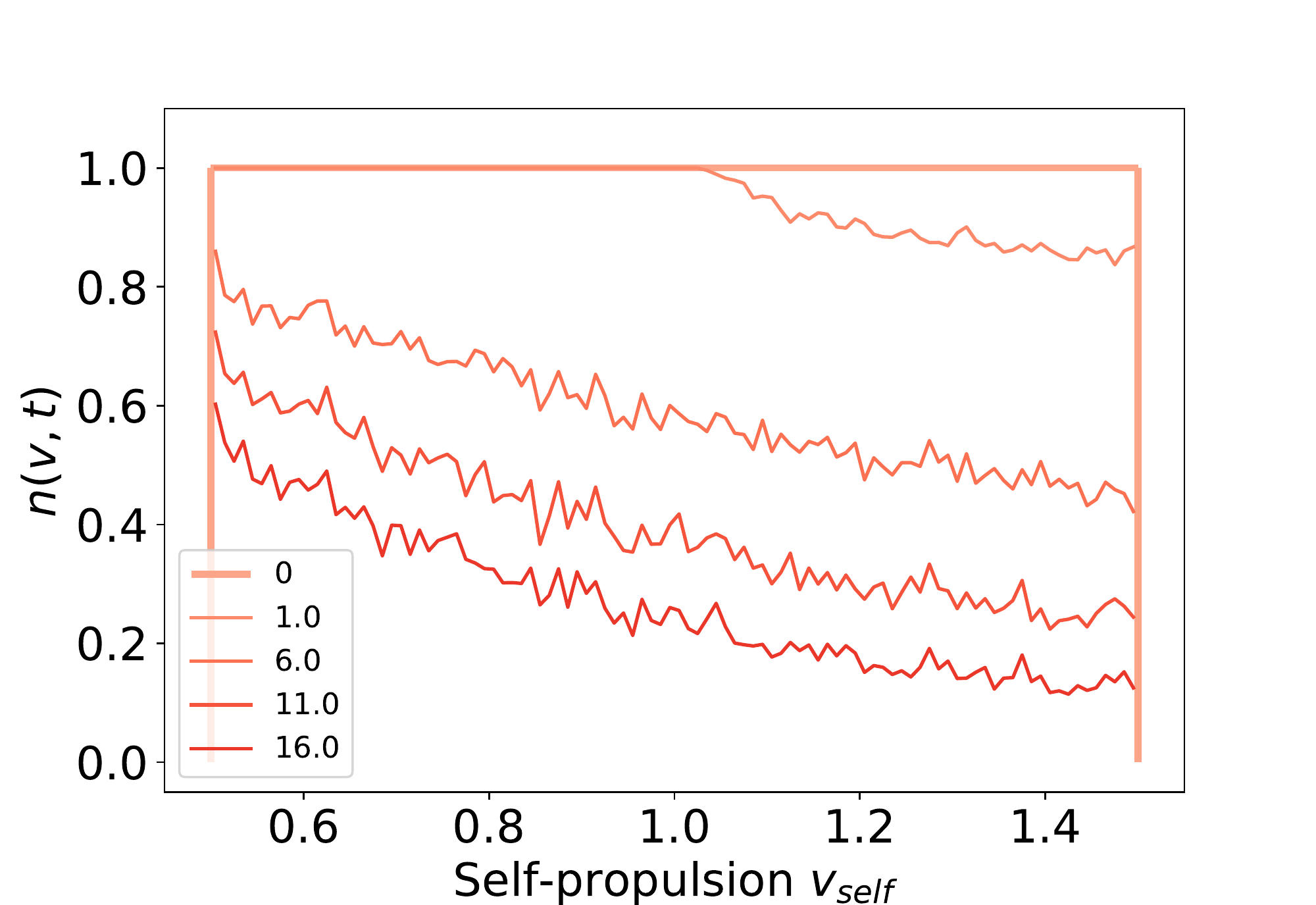} 
\caption{ Mixture of particles with different velocities.  
  Time evolution of $n(v,t)$
  at \matteo{five} representative times, the radius of the chamber is $R=1$ and
  rotational diffusion $D_r=1$. The initial
  distribution that is uniform in the interval $v\in
  [0.5,1.5]$. Faster particles with $v > 1$ escape very early from the
  circular vessel. 
      }
\label{fig:fig3}      
\end{figure}

\section{Designing simple sorting devices}

\begin{figure}[!t]
\centering
\includegraphics[width=.5\textwidth]{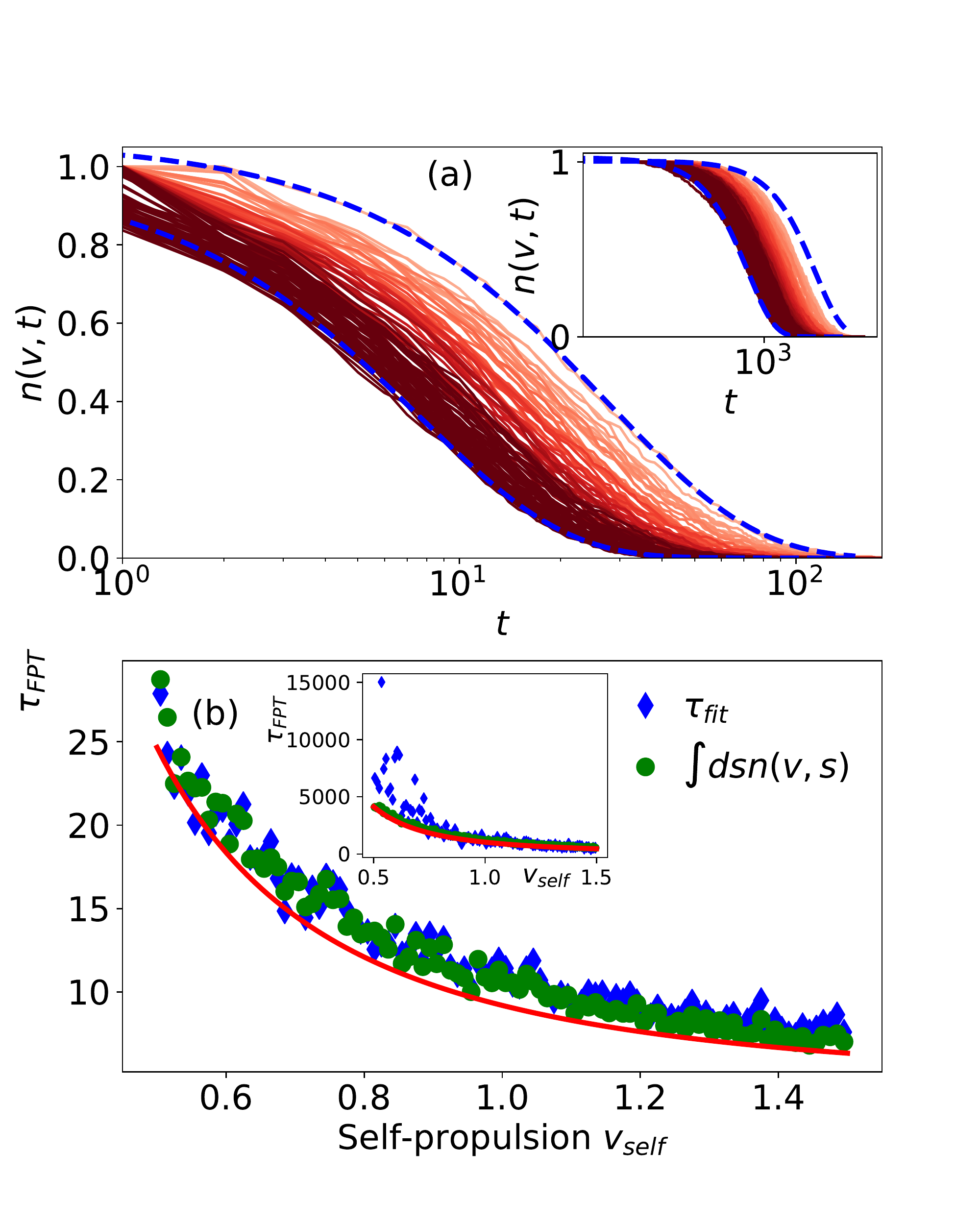} 
\caption{Decay of the number density for different $v_{self}$.
   (a) Time evolution of $n(v,t)$, velocity decreases from $1.5$
  (dark red) to $0.5$ (light red) The decay of $n(v,t)$ is well
  captured by an exponential curve (dashed blue lines are fit with
  $n_{fit}(v,t) = A e^{-t / \tau(v)}$ ). Inset: same observable for
  $R=10$.   \matteo{(b) The decay time $\tau(v)$ computed by fitting to an
  exponential decay (blue diamonds) and using Eq. (8) with $R=1$. 
  Solid red curve is the empirical function $f(x)$ (in the inset $R=10$).}}
%
\label{fig:fig4}      
\end{figure}

In this section we explore the possibility of tuning the geometrical
properties of the confining structure for sorting particles of
different velocities. The advantage of this approach relies on the
fact that (i) we do not have to introduce any external field, (ii) the
geometry is extremely simple and easily realised in microfluidics, as
compared to~ \cite{Galajda07}, (iii) the only parameter we have to
tune is the radius $R$.

According to the results of the previous sections, $\tau_{FPT}$
shows a crossover around $R \ell^{-1}=1$ between two regimes
characterized by different scaling laws.
To make it explicit, we rewrite here
Eq.~(\ref{empir}) in terms of the motility parameters for the mean exit
time: 
\BEQ \label{empi2} \tau_{FPT} = \alpha \frac{R^2}{v_{self}^2
  \tau_{pers}} + \beta \tau_{pers}.  
  \EEQ 
  It is immediately understood
that changing $v_{self}$ at constant $\tau_{pers}$ leads to a monotone
behavior, with a crossover at $v_{self} \approx v_{self}^*$, with
$v_{self}^*= \sqrt{\alpha/\beta} R/\tau_{pers}$, from a diffusive
regime with slow escape to the active regime with fast escape. On the
contrary, changing $\tau_{pers}$ at constant $v_{self}$ produces a
{\em non-monotone} behavior with an optimal escape at $\tau_{pers}
\approx \tau_{pers}^*$, with 
$\tau_{pers}^* = \sqrt{\alpha/\beta} R/v_{self}$ 
and slower escape both form smaller and larger
values of $\tau_{pers}$. In both cases we foresee applications in
sorting problems, with different ways of use.

\subsection{Different propulsion velocities}

As a model system, we start by considering a gas composed by particles
with different self-propulsion velocity.  The velocity is extracted by
a uniform distribution.  We consider a system composed by $N=5 \times
10^{4}$ Active Brownian particles in a circular chamber of size $R =
1$ where a small slit of size $\delta=\matteo{2 \sigma}=1$ allows particles to
escape.  The system is characterized by the initial distribution of
self-propulsion velocities $\rho(v) = \frac{1}{v_{max} - v_{min}}
\left[ \vartheta(v - v_{min}) - \vartheta(v - v_{max}) \right]$, with
$\vartheta(x)$ the Heaviside step function. Here $v_{max} = 1.5$ and
$v_{min} = 0.5$.  All particles have the same rotational diffusion
constant $D_{r} = \tau_{pers}^{-1} = 1$. 
In this way, each particle $i$ has
its own persistence length $\ell_i = v_{self}^i$. According to
previous results we expect a crossover at $v_{self}^* \approx 1$: faster
  particles escape within a time of order $1$ on average, slower
  particles escape in a much slower time, sensitive upon $v_{self}$.

\begin{figure}[!t]
\centering
\includegraphics[width=.5\textwidth]{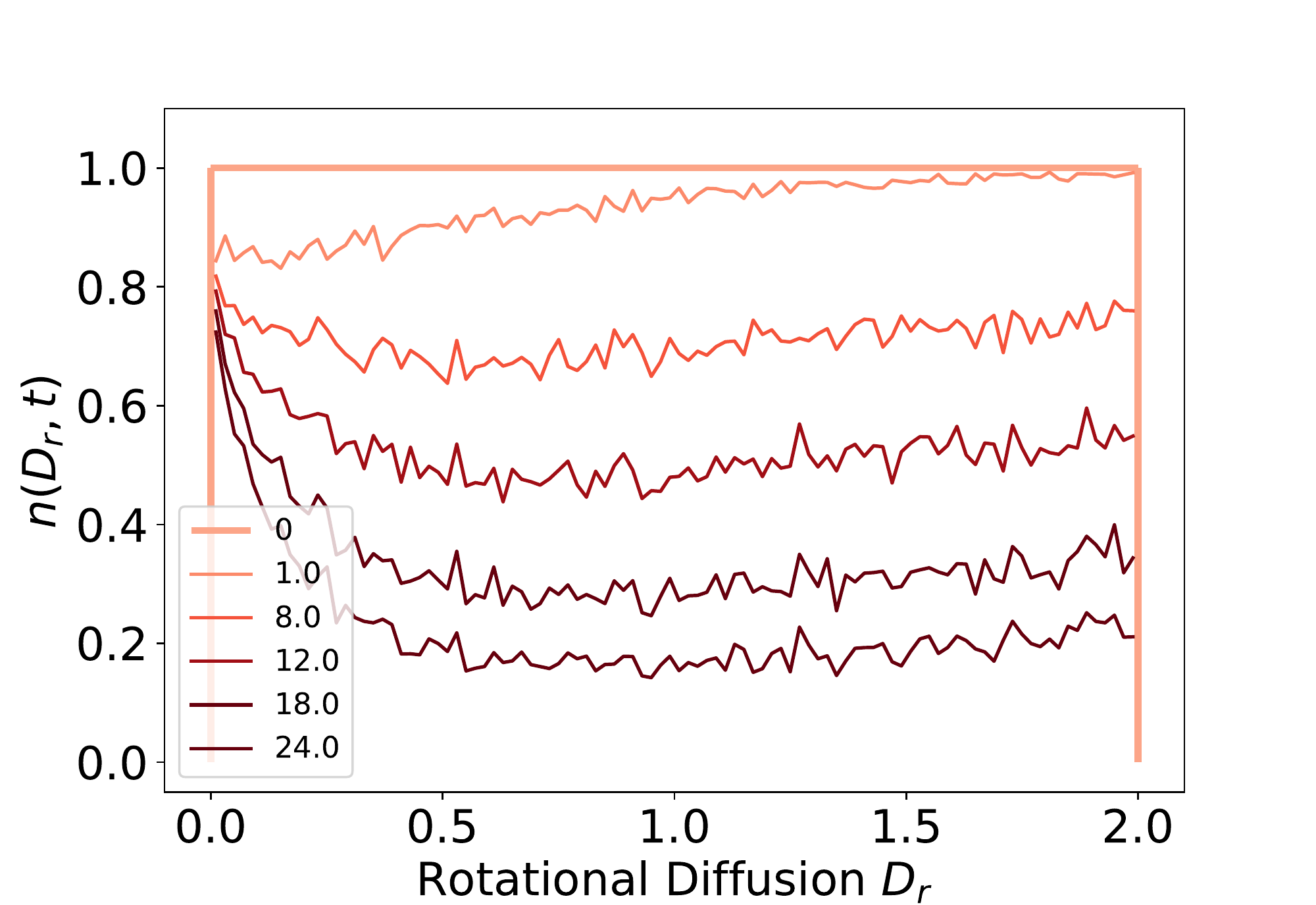} 
\caption{ Mixture of particles with different persistence times.
  Time evolution of $n(D_r,t)$ at six representative times, the radius 
  of the chamber is $R=1$. The initial distribution $n(D_r,0)$ is uniform 
  in the interval $D_r \in [0,2]$.
\label{fig:fig5}      }
\end{figure}

We are interested in the time evolution of the number of particles
with velocity $v$ that falls into the interval $v \in [v,v+dv] $, with
$dv=(v_{max} - v_{min}) / \mathcal{N}$, with $\mathcal{N}=100$.  

\matteo{In Fig. (\ref{fig:fig3})}
the time-evolution of $n(v,t)$ is shown as a function of $v$
for $5$ time steps.  Dashed red curve indicates the
initial uniform configuration. The second distribution is taken at
$t=1$, as soon as the first particles have found the exit.  As one
can see, the only particles escaped are those with $v_{self} >
v_{self}^*$.  With increasing time, also particles with
$v_{self}<v_{self}^*$ start to escape from the chamber, with times
that depend upon $v_{self}$. 

An alternative - and informative - way of seeing these results is
studying the relaxation dynamics towards zero of $n(v,t)$ as a
function of time. We recall that $n(v,t)$ at a given $v=v_{self}$ is
the survival probability in this problem, which is the complementary
cumulative of the probability density function $p(v,s)$ of having
first exit at time $s$: 
\BEQ n(v,t)=\int_t^{\infty} ds p(v,s), 
\EEQ 
or equivalently 
\BEQ p(v,s) = -\frac{d n(v,t)}{dt}. \label{surv} 
\EEQ
The results are reported in Fig. (\ref{fig:fig4}). In panel (a),
$n(v,t)$ is shown for different values of self-propulsion velocities,
increasing from light to dark red. 
\matteo{We notice that a simple exponential relaxation characterized by
a single velocity-dependent time-scale $\tau(v)$ might capture the behavior 
of $n(v,t)$ as a function of time.}
We show the results of exponential fits of the decay
$n_{fit}(v,t) = A \exp{ (-t / \tau_{fit} (v))}$, with $A \sim
1$. Dashed blue curve in panel (a) are two representative fits for
$v_{min}=0.5$ and $v_{max}=1.5$. 
 In the inset of the same panel, the decay of
$n(v,t)$ for a larger chamber is also reported, for comparison. In
this case, we have increased the radius of the circular confinement of
an order of magnitude, i. e., $R = 10$, making the threshold velocity
$v_{self}^* \approx 10$. All particles have now a persistence length
much smaller than the size of the chamber and are in the diffusive regime, with slow escape.
%
%
%

\matteo{Moreover, i}n view of Eq.~(\ref{surv}), the knowledge of $n(v,t)$ gives us immediate
access to the average mean first exit time:
\begin{multline} \label{fptave}
\tau_{FPT}(v) =
\int_0^{\infty} ds s p(v,s) = \\-\int_0^{\infty} ds \, s\frac{d n(v,s)}{ds} 
= \int_0^{\infty} \matteo{ds} \, n(v,s).
\end{multline}
\matteo{It is worth noting that, in the case of a purely exponential relaxation, i. e.,  $n(v,t)=e^{-t/\tau_{fit}(v)}$,
$\tau_{FPT}(v)$ coincides with $\tau_{fit}$.} 
The behavior of $\tau_{FPT}$ as a function of $v_{self}$ is shown in
panel (b). 
\matteo{As one can appreciate, $\tau_{FPT}$ and $\tau_{fit}$ are in a nice agreement 
for $R=1$. The solid curve is Eq. (\ref{empi2}) with the values of the parameters $\alpha$ and $\beta$
previously fitted.}
In the experiment with $R=1$ (main plot) the crossover can
be appreciated from small to large velocities where $\tau_{fit}$
decays and then reaches a plateau.  When $R=10$, on the contrary, the
plateau is not reached and the $1/v_{self}^2$ scaling can be
appreciated, see Eq.~(\ref{empi2}). 

\begin{figure}[!t]
\centering
\includegraphics[width=.5\textwidth]{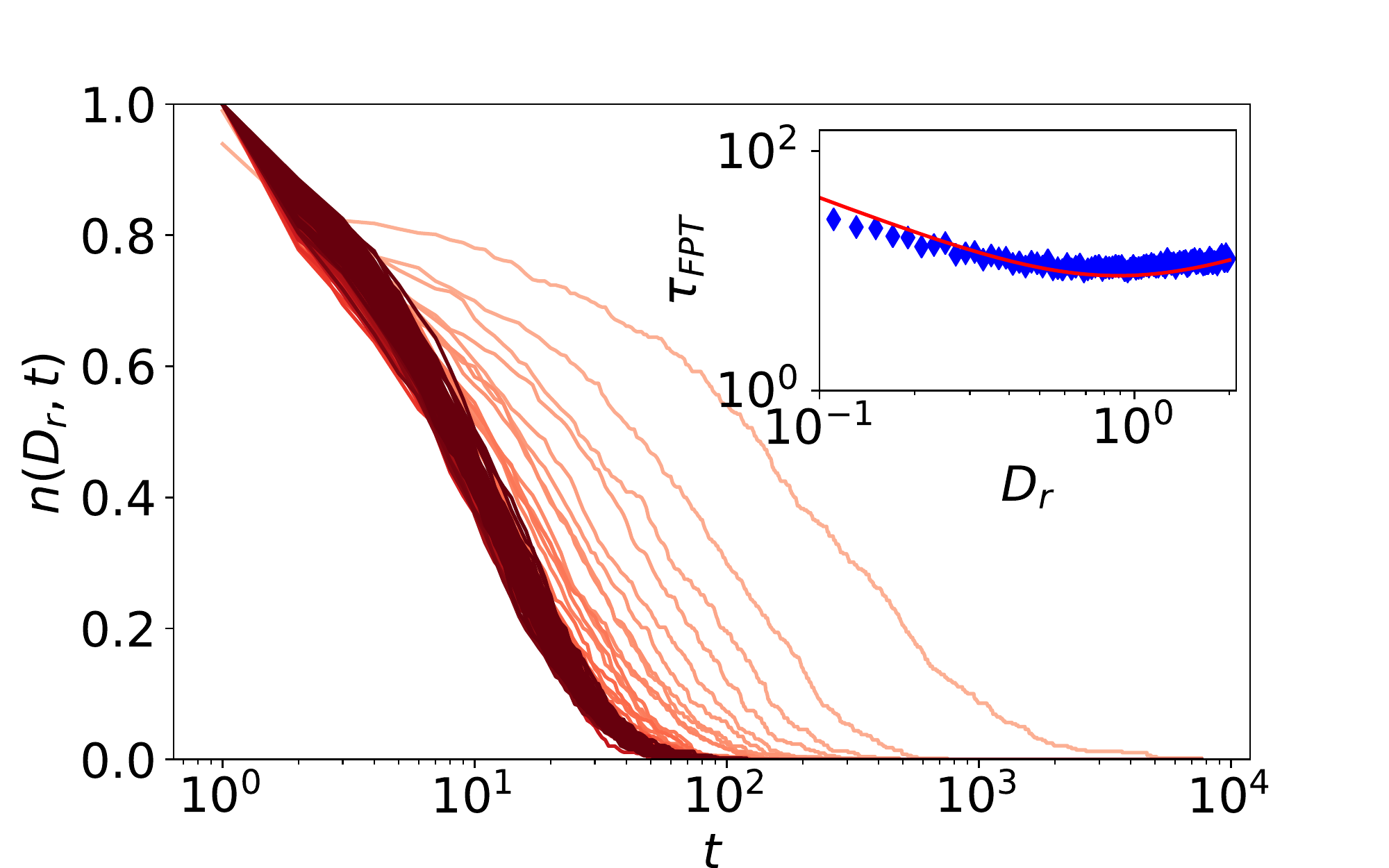} 
\caption{ Decay of the number density for different $D_r$.  Time
  evolution of $n(D_r,t)$, the rotational diffusion $D_r$ decreases
  from $2$ (dark red) to $0$ (light red).  The inset shows the decay
  time $\tau$ as a function of $D_r$. 
  \matteo{Solid red curve is $f(x)$ with $\alpha$ and $\beta$
  previously fitted from $\tau_{FPT}$.}}
\label{fig:fig6}      
\end{figure}

\subsection{Different persistence times}

In this section we tune the persistence length $\ell$ by varying the
persistence time $\tau_{pers}$ - which is $1/D_r$ for the case
considered here of Active Brownian particles - and maintaining fixed
the self-propulsion velocity $v_{self}$.  We investigate the behavior
of a sample composed by the same number of particles of the previous
case with same self-propulsion velocity, i. e., $v_{self} = 1$, and
characterized by an initial distribution of rotational diffusion
$\rho(D_r) = \frac{1}{D_{r,max} - D_{r,min}} \left[ \vartheta(D_r -
  D_{r,min}) - \vartheta(D_r - D_{r,max}) \right]$, with $D_{r,max}=2$
and $D_{r,min}=0$. The results are shown in
Fig. (\ref{fig:fig5}). Note that, in this case, particles with $D_r=0$ remain
trapped unless they starts with the right direction.

In Fig. (\ref{fig:fig6}) the typical behavior of $n(D_r,t)$ as a
function of time is shown. In this case we observe for the very low
values of $D_r$ (high activity) that a single exponential decay does
not reproduce the decay of $n(D_r,t)$: interestingly, the survival
probability seems to follow a first decrease to a plateau followed by
a second final decay.
%
%
%

Again we measure $\tau_{FPT}$ from Eq.~(\ref{fptave}) and plot it in
the inset of Fig.~\ref{fig:fig6}. According with the scaling of
$\tau_{FPT}$ we discussed in section~\ref{results}, $\tau_{FPT}$
matches two asymptotic regimes, the first one, for small $D_r$ values
and thus large persistence length, is $\tau_{FPT} \sim D_r^{-1}$. The
second one is typical of the diffusive regime, meaning that $\tau_{FPT}
\sim D_r$. In the inset of the same figure we report $\tau_{FPT}$ as a
function of $D_r$, with superimposed Eq.~(\ref{empi2}).  

\begin{figure}[!t]
\centering
\includegraphics[width=.5\textwidth]{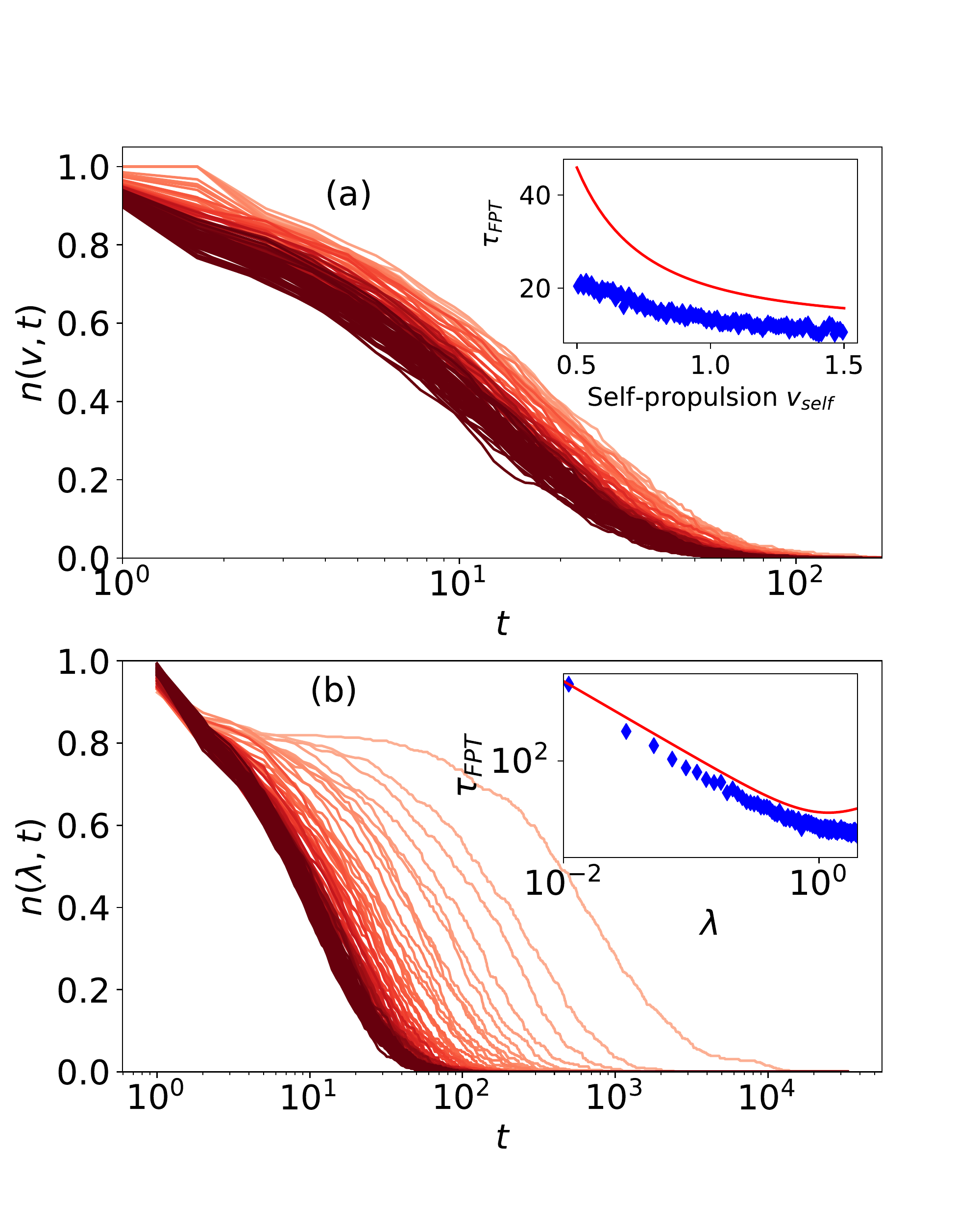} 
\caption{ Sorting of Run-and-Tumble particles.  (a) Time
  evolution of $n(v,t)$ as a function of $t$ with $v\in[0.5,1]$. The
  inset show the behavior of $\tau$. 
  The dashed red curve is $f(x)$. (b) Time evolution of $n(\lambda,t)$ as a function of $t$
  with $ \lambda \in [0,2]$.  In the inset we report $\tau$ as a
  function $\lambda$.  }
\label{fig:fig7}      
\end{figure}

\subsection{Behavior of Run-and-Tumble particles}

While the results in the previous sections have been obtained for
Active Brownian particles, we have performed the same numerical
simulations in the case of Run-and-Tumble dynamics, again for mixtures
of particles with different propulsion velocities \matteo{$v_{self} \in [0.5,1.5]$}
and persistence times $\tau_{pers}=1/\lambda$ \matteo{with $\lambda \in [0.01,2] $.} 
  The results, shown in Fig. (\ref{fig:fig7}) (a) and (b)
respectively, are qualitatively the same of the previous cases: the
relaxation dynamics of $n(v,t)$ and $n(\lambda,t)$ are well described
by an exponential decay, unless the persistence time is very large
($\lambda \ll 1$). We notice that the non-exponential decay occurs for
a larger range of persistence time with respect to the Active Brownian
case (compare Fig. 7-(b) and Fig. 6): in our opinion this observation
corroborates a correlation between the {\em two-steps} decay and
ballistic behavior, which suggests that typical exit trajectories at
large $\tau_{pers}$ are composed of two separate processes: first, the
particle reach the boundary and second the particles finds the exit
remaining close to the boundary .  The behavior of $\tau_{FPT}$ is
also reported in the insets of the two plots in Fig. (\ref{fig:fig7}).



\section{Discussion and Conclusions} \label{Discussion}

In this work we have proposed a numerical study of the narrow escape
problem \matteo{of} active particles in circular domains. We compared two
paradigmatic models of active motions that are Active Brownian
dynamics \cite{bialke2015active}, suitable for reproducing the
trajectories of smooth swimmers, and Run-and-Tumble dynamics, that,
for instance, well captures the morphological properties of {\it
  E. coli} trajectories \cite{berg2008coli}. We showed that in both
\matteo{dynamics} 
$\tau_{FPT}$ turns to be bounded by two limiting asymptotic
\matteo{regimes. The two regimes result from the competition between
the two characteristic length of the system: the persistence length $\ell$
and the radius $R$ of the container. When the persistence length is much smaller
than the radius, i. e., $\ell / R \ll 1$, the active particles behaves as 
a Brownian random walker and $\tau_{FPT}$ diverges as the effective diffusivity
of the random walk goes to zero. In the opposite limit, the persistence length
is much larger than the size of the chamber, i. e., $\ell / R \gg 1$.  
The escaping dynamics is thus dominated by the ballistic regime
and $\tau_{FPT}$ grows linearly with $\tau_{pers}$.}
\matteo{We have introduced an empirical scaling function $f(x)$, with
$x=\ell^2 R^{-2}$, that smoothly connects these two asymptotic regimes.}


We thus explored the possibility to take advantage of the crossover in
the escape time between the active and the diffusive regime for
sorting particles of different velocities.  We obtained that,
considering a gas of Active Brownian particles of different velocities
and same rotational diffusion constant, by tuning the size of the
chamber faster particles can be separated by the others.  The same
technique can be employed for demixing particles of different
rotational diffusion.  
We showed that the same is
true also in the case of Run-and-Tumble dynamics.

\matteo{It is worth noting that the} sorting mechanism \matteo{explored here} is dynamical 
and \matteo{it} works only
on a finite time scale. In particular, waiting sufficiently long time,
all particles escaped from the confining structure.  However, in
practical situation, the typical experimental time scale can be tuned
for sorting microswimmers of different velocities tuning just the
typical size of the confining structures, i. e., without introducing
any external potential or complicated microsctructure. This could be
useful in assistant reproductive technologies, where the challenge
consists in maximize motile sperm concentration, sperm volume and lifetime.
Usually, sperms are selected based on their motility. Our findings
suggests that it could be done without introducing density gradient
centrifugation \cite{nosrati2017microfluidics,koh2015study}.


\section*{Acknowledgments} 
The research leading to these results has received funding from Regione Lazio, Grant Prot. n. 
85-2017-15257 ("Progetti di Gruppi di Ricerca - Legge 13/2008 - art. 4").
\bibliography{mpbib}
\bibliographystyle{apsrev}

\end{document}